\documentclass[%
 reprint,
 superscriptaddress,
 amsmath,amssymb,
 aps,
 prl,
]{revtex4-2}
\usepackage{graphicx}% Include figure files
\usepackage{dcolumn}% Align table columns on decimal point
\usepackage{bm}% bold math
\usepackage{siunitx}
\usepackage{xcolor}
\usepackage{comment}
\usepackage{adjustbox}
\usepackage[utf8x]{inputenc}
\usepackage{ucs}
\usepackage[none]{hyphenat}
\usepackage{multirow}
\usepackage[normalem]{ulem}
\usepackage{subfigure}

\begin{document}
% \linenumbers
\title{ 
Search for Light Dark Matter with 259 Days of Data in PandaX-4T 
} 
% !TEX root = ../main.
%updated on 2025/06/09

\def\tdli{State Key Laboratory of Dark Matter Physics, Key Laboratory for Particle Astrophysics and Cosmology (MoE), Shanghai Key Laboratory for Particle Physics and Cosmology, Tsung-Dao Lee Institute \& School of Physics and Astronomy, Shanghai Jiao Tong University, Shanghai 201210, China}
\def\sjtuphys{State Key Laboratory of Dark Matter Physics, Key Laboratory for Particle Astrophysics and Cosmology (MoE), Shanghai Key Laboratory for Particle Physics and Cosmology, School of Physics and Astronomy, Shanghai Jiao Tong University, Shanghai 200240, China}
\def\newcorner{New Cornerstone Science Laboratory, Tsung-Dao Lee Institute, Shanghai Jiao Tong University, Shanghai 201210, China}
\def\MESJTU{School of Mechanical Engineering, Shanghai Jiao Tong University, Shanghai 200240, China}
\def\SPEIT{SJTU Paris Elite Institute of Technology, Shanghai Jiao Tong University, Shanghai 200240, China}
\def\SJTUSC{Shanghai Jiao Tong University Sichuan Research Institute, Chengdu 610213, China}

\def\BUAA{School of Physics, Beihang University, Beijing 102206, China}
\def\BUAACenter{Peng Huanwu Collaborative Center for Research and Education, Beihang University, Beijing 100191, China}
\def\BUAALab{International Research Center for Nuclei and Particles in the Cosmos \& Beijing Key Laboratory of Advanced Nuclear Materials and Physics, Beihang University, Beijing 100191, China}
\def\SCNT{Southern Center for Nuclear-Science Theory (SCNT), Institute of Modern Physics, Chinese Academy of Sciences, Huizhou 516000, China}

\def\USTClab{State Key Laboratory of Particle Detection and Electronics, University of Science and Technology of China, Hefei 230026, China}
\def\USTCdep{Department of Modern Physics, University of Science and Technology of China, Hefei 230026, China}

\def\YaLongSD{Yalong River Hydropower Development Company, Ltd., 288 Shuanglin Road, Chengdu 610051, China}
\def\scKeyLab{Jinping Deep Underground Frontier Science and Dark Matter Key Laboratory of Sichuan Province, Liangshan 615000, China}

\def\pku{School of Physics, Peking University, Beijing 100871, China}
\def\CHEPpku{Center for High Energy Physics, Peking University, Beijing 100871, China}

\def\SDUdep{Research Center for Particle Science and Technology, Institute of Frontier and Interdisciplinary Science, Shandong University, Qingdao 266237, China}
\def\SDUlab{Key Laboratory of Particle Physics and Particle Irradiation of Ministry of Education, Shandong University, Qingdao 266237, China}

\def\UMD{Department of Physics, University of Maryland, College Park, Maryland 20742, USA}

\def\SYU{School of Physics, Sun Yat-Sen University, Guangzhou 510275, China}
\def\SYUSFI{Sino-French Institute of Nuclear Engineering and Technology, Sun Yat-Sen University, Zhuhai 519082, China}
\def\SYUzhuhai{School of Physics and Astronomy, Sun Yat-Sen University, Zhuhai 519082, China}
\def\SYUshenzhen{School of Science, Sun Yat-Sen University, Shenzhen 518107, China}

\def\NKU{School of Physics, Nankai University, Tianjin 300071, China}
\def\YTU{Department of Physics, Yantai University, Yantai 264005, China}
\def\FDU{Key Laboratory of Nuclear Physics and Ion-beam Application (MOE), Institute of Modern Physics, Fudan University, Shanghai 200433, China}
\def\CDUT{College of Nuclear Technology and Automation Engineering, Chengdu University of Technology, Chengdu 610059, China}

\affiliation{\tdli}
\author{Minzhen Zhang}\affiliation{\tdli}
\author{Zihao Bo}\affiliation{\sjtuphys}
\author{Wei Chen}\affiliation{\sjtuphys}
\author{Xun Chen}\affiliation{\tdli}\affiliation{\SJTUSC}\affiliation{\scKeyLab}
\author{Yunhua Chen}\affiliation{\YaLongSD}\affiliation{\scKeyLab}
\author{Chen Cheng}\affiliation{\BUAA}
\author{Xiangyi Cui}\affiliation{\tdli}
\author{Manna Deng}\affiliation{\SYUSFI}
\author{Yingjie Fan}\affiliation{\YTU}
\author{Deqing Fang}\affiliation{\FDU}
% ...
\collaboration{PandaX Collaboration}
\noaffiliation

\date{\today}% It is always \today, today,

\begin{abstract}
  
We present a search for light dark matter particles through their interactions with atomic electrons and nucleons, utilizing PandaX-4T data with an effective exposure of 1.04~tonne$\cdot$year for ionization-only data and 1.20~tonne$\cdot$year for paired data. 
Our analysis focuses on the energy range (efficiency$>$0.01) of approximately 0.33 to 3~keV for nuclear recoils, and from 0.04 to 0.39~keV for electronic recoils. 
We establish the most stringent constraints on spin-independent dark matter-nucleon interactions within a mass range of 2.5-5.0~GeV/$c^2$, spin-dependent neutron-only interactions within 2.0-5.3~GeV/$c^2$, and spin-dependent proton-only interactions within 2.0-3.8~GeV/$c^2$. 
Their corresponding limits at 3\,GeV/$c^2$ are $1.1 \times 10^{-43}$, $1.6 \times 10^{-38}$, and $5.6 \times 10^{-37}$\,cm$^2$, respectively.
Additionally, our results improve the upper limits on the dark matter-electron scattering cross section by a factor of 1.5 and 9.3 for heavy and light mediator scenarios, respectively, within 50~MeV/$c^2$ to 10~GeV/$c^2$, compared with previous best results.
The improved limits reach 1.5$\times$10$^{-41}$ and 5.6$\times$10$^{-37}$cm$^2$ at 200\,MeV/$c^2$ for the heavy and light mediators, respectively.

% Such light dark matter particles with mass in [xxx, xxx]\,MeV \textcolor{red}{blabla (main theorectical model feature)}, thus, exhibiting a diurnal modulation of the data distribution due to Earth's rotation.
% A signal significance of \textcolor{red}{XXX} is obtained for such light dark matter model.
% The best-fit gives a dark matter-matter cross section of \textcolor{red}{XXX$\pm$XXX\,cm$^2$} and a mean dark matter particle direction of \textcolor{red}{XXX}.

\end{abstract}

\maketitle

% quick functions
\newcommand{\mwba}[1]{\textcolor{violet}{#1}}
\newcommand{\mwbd}[1]{\textcolor{violet}{\sout{#1}}}

% Introduction about light dark matter
A Large number of astrophysical observations have supported the common existence of dark matter (DM) in our Universe~\cite{Zwicky_1933gu, Rubin_1970zza, Umetsu_2011, Clowe_2006, Planck2016}, but the nature of DM is still unknown.
Cold DM particles, such as weakly interacting massive particles (WIMPs)~\cite{susy, dm_candidate}, are considered one of the most promising candidates.
The most probable mass range of such cold DM particles is between about several GeV/$c^2$ and approximately 10\,TeV$/c^2$. 
Decades-long searches for WIMPs have not yielded conclusive detections yet. However, they led to the development of various ultrasensitive detection techniques~\cite{dm_direct_detection_review}.
Recent astrophysical observations of the stellar cusps and dwarf galaxies reveal a deviation from the standard cold DM~\cite{dm_candidate}.
They promote studies of the light DM (LDM) particles which have masses approximately from keV/$c^2$ to sub-GeV/$c^2$. There are many LDM candidates, including sterile neutrinos~\cite{Boyarsky:2009ix}, axions and axionlike particles~\cite{Marsh:2015xka}, hidden sector particles~\cite{Feng:2011ik}, dark photons~\cite{Feldstein:2013kka}, superWIMPs~\cite{Feng:2003uy}, and DM from primordial black hole evaporation~\cite{Masina:2020xhk}, etc. LDM, produced via freeze-in, freeze-out, or nonthermal mechanisms, typically interacts weakly with the standard model particles, addressing small-scale structure issues and offering diverse detection signatures in astrophysical and cosmological observations~\cite{Arcadi:2017kky}. 
LDM interactions may evade detection owing to their extremely low energy deposition.
In this work, we use the low-energy data ($\sim$keV and sub-keV regions) from the PandaX-4T experiment to search for the LDM particles through their scatters off the target nuclei or atomic electrons.

% PandaX-4T
The PandaX-4T~\cite{PandaX-4T:2021bab, bo2025dark}, located at the China Jinping Underground Laboratory (CJPL)~\cite{kang2010status, li2015second}, is a deep-underground multitonne liquid xenon detector designed to search for ultrarare events from dark matter particles, Majorana neutrinos, and astrophysical neutrinos. 
It utilizes a dual-phase cylindrical time projection chamber (TPC) with a diameter of $\sim$1.2 m and a height of $\sim$1.2 m. 
The detector uses approximately 5.6 metric tons of liquid xenon as the target material, with around 3.7 metric tons within the sensitive volume.
It is equipped with 169 and 199 Hamamatsu R11410-23 photomultipliers (PMTs), which are highly sensitive to ultraviolet light, positioned at the top and bottom of the TPC, respectively. 
The average detection efficiency of the PandaX-4T TPC for the prompt scintillation light signals ($S1$) from an energy deposition is approximately 10\%. 
The remaining detectable energy is carried by ionized electrons, which drift to the gaseous layer and amplified, producing a secondary scintillation signal referred to as $S2$. 
The detection efficiency for these ionized electrons typically reaches $\sim$70\% in average, with the efficiency losses of about 25\% attributed to attachment to impurities during the electron drifting, and of about 10\% attributed to incomplete extraction of the drifted electrons at the liquid-gas interface. 
Further details regarding the experimental setup can be found in Refs.~\cite{luo2024signal, He:2021sbc, Yang:2021hnn, Zhao:2020vxh, Yan:2021cxp, PandaX-4T:2021lbm}.

\begin{figure*}[htp]
    \centering
    \includegraphics[width=0.85\textwidth]{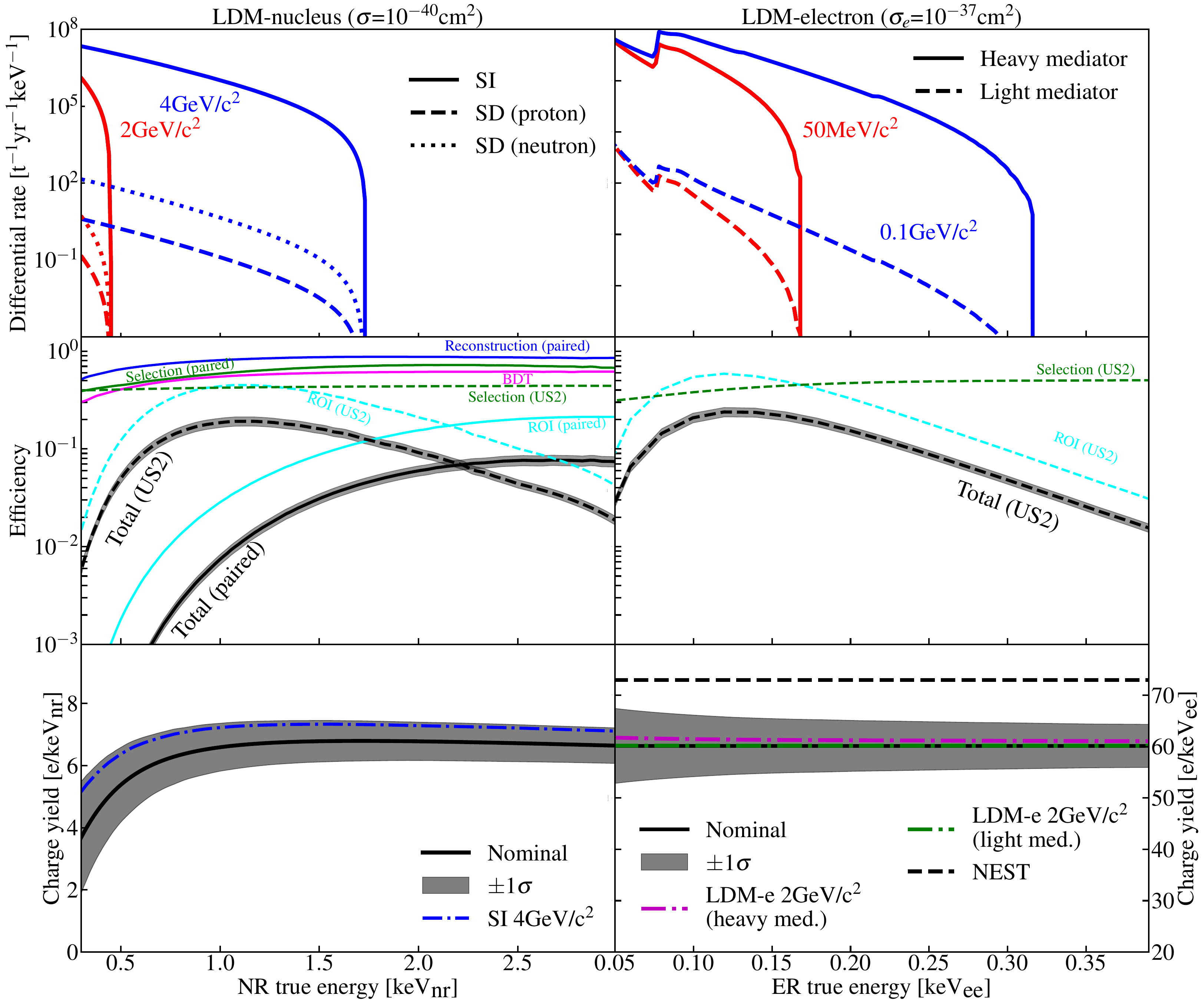}
    \caption{
    Top: the differential rates (in unit of events per metric ton per year per keV) of the DM scatterings off the target.
    The left shows the spectra for the LDM-nucleus scattering with LDM masses of 2\,GeV (red) and 4\,GeV (blue), assuming an LDM-nucleon cross section of 10$^{-40}$\,cm$^2$. 
    The spectra for the SI, proton-only SD, and neutron-only SD LDMs are shown in solid, dashed, and dotted lines, respectively.
    The right panel shows the spectra for LDM-electron scatterings with LDM masses of 50\,MeV/$c^2$ (red) and 0.1\,GeV (blue), assuming an LDM-electron cross section of 10$^{-37}$\,cm$^2$.
    The solid and dashed lines represent the spectra for heavy and light mediator cases, respectively.
    All spectra are without efficiency applied.
    Middle: all components of the selection efficiency for the paired (solid) and US2 (dashed) data: green, data quality selection; blue, signal reconstruction; cyan, ROI; magenta, BDT; black, total.
    The gray shaded regions indicate $\pm$1$\sigma$ uncertainties.
    The efficiencies for the paired data are not shown in the right panel since they are not used in search of LDM-electron scattering.
    The efficiencies as a function of NR and ER energies are estimated based on the nominal charge yields shown in the bottom panels.
    Bottom: the charge (black) yields as a function of NR and ER energies are shown in the left and right panel, respectively.
    The solid lines show the nominal model with the shaded regions representing the $\pm$1$\sigma$ uncertainties.
    The black dashed lines in the right panel give the alternative yield model (P4-NEST) used in this work. 
    The colored lines (blue, magenta, and green) give the best-fit charge yields in the LDM interpretations with different masses and model assumptions.
    % \textcolor{red}{(Todo: 1) Update the efficiencies; 2) Update the Yield models.)}
    No significant deviation from the nominal models is observed.
    }
\label{fig:signal_and_efficiency}
\end{figure*}
 
% Physical Model
In DM direct detection experiments, benchmark searches typically assume that DM particles interact with the target nuclei via spin-independent (SI) interactions. 
This assumption results in a total cross section for DM-nucleus interactions that scales with the target's atomic mass due to the coherent scattering effect.
%However, there may also be an axial-vector contribution to the DM-nucleus cross section, introducing spin dependence to the interaction.
However, there may also be a spin-dependent (SD) contribution to the DM-nucleus cross section, like axial-vector mediated interaction, etc.
Only nuclear isotopes with nonzero spin can participate in the SD scattering processes.
Generally, the differential rate of DM-nucleus interactions can be expressed as an integral over the DM velocity $v$:
\begin{equation}
\frac{dR}{dE_{\textrm{NR}}} = \frac{\rho}{2m_\chi \mu_N^2} \int dv \frac{f(v)}{v} \left(
\sigma^{\textrm{SI}}_{0, N} F^2_{\textrm{SI}} + \sigma^{\textrm{SD}}_{0, N} F^2_{\textrm{SD}}
\right),
\label{eq:nr}
\end{equation}
where $\rho$, $m_\chi$, and $\mu_N$ are the local DM density, the mass of the DM particle, and the reduced mass between DM and the target nucleus, respectively.
The $f(v)$ represents the Maxwellian velocity distribution of the local DM particles traversing the detector.
The $\sigma^{\textrm{SI}}_{0, N}$ and $\sigma^{\textrm{SD}}_{0, N}$ denote the energy-independent SI and SD DM-nucleus cross sections, respectively, while the $F^2_{\textrm{SI}}$ and $F^2_{\textrm{SD}}$ are the corresponding energy-dependent nuclear form factors.
$F_{\textrm{SI}}$ was taken from Ref.~\cite{lewin1996review}.

In SI DM searches, $\sigma^{\textrm{SI}}_{0, N}$ is simplified to $\bar{\sigma} \mu_N^2 A^2 / \mu_n^2$, where $\mu_n$ is the reduced mass for DM-nucleon and $A$ is the target mass number.
The effective DM-nucleon cross section $\bar{\sigma}$ is conventionally compared across experiments. 
For a proton-only SD interaction, the term $\sigma^{\textrm{SD}}_{0, N} F^2_{\textrm{SD}}$ can be expressed as $\sigma_p \frac{4\pi}{3(2J+1)} \frac{\mu_N^2}{\mu_p^2} S_A$, where $J$ is the nuclear spin, $S_A$ is the energy-dependent structure function~\cite{Klos:2013rwa}, and $\sigma_p$ is the DM-proton cross section.
In the case of neutron-only interaction, $p$ can be replaced by $n$. 
In principle, the SD interaction can be a combination of proton-only and neutron-only interactions, depending on the assumptions made regarding the interaction, which can be effectively decomposed into various operators~\cite{eft_dm}. 
For simplicity, this work focuses on the proton-only and neutron-only cases to compare with constraints presented in the literature. Given the fact that the xenon nucleus has an even number of protons, the spin of a xenon nucleus is dominated by the unpaired neutron due to the spin pairing effects and the event rate of the neutron-only interaction is significantly larger than the proton-only case even with the same DM-nucleon cross section.

The interaction between DM and atomic shell electrons is also of great interest, due to its lower energy required to produce detectable quanta in detector.
This interaction channel is particularly sensitive to LDM that is too light to produce sufficient deposit energy through nuclear recoils (NRs) but heavy enough to scatter off atomic electrons, depositing a considerable amount of electronic recoil (ER) energy.
The differential rate for the LDM-electron scattering can be expressed as integral over $(n,l)$ shell contributions~\cite{essig2017new}:
\begin{equation}
    \frac{dR}{dE_{\textrm{ER}}} = \frac{\rho \bar{\sigma}_e}{8m_\chi \mu^2_e}\sum_{nl} \int qdq \left| f^{nl}_{\textrm{ion}} (k^\prime, q) \right|^2 \left| F_e\right|^2 \eta(v_{\textrm{min}}),
    \label{eq:er}
\end{equation}
where $\bar{\sigma}_e$ is the cross section of LDM scattering off free electron at fixed momentum transfer $q=\alpha m_e$, with $\alpha$ the fine-structure constant and $m_e$ the mass of electron.
$\eta(v_{\textrm{min}})$ is the inverse mean speed, $\mu_e$ is the reduced LDM-electron mass, and $F_e$ is the momentum transfer-dependent LDM-electron form factor.
In this analysis, we discuss the LDM-electron scatterings with heavy ($F_e=1$) and light mediators ($F_e \propto 1/q^2$).
The ionization form factor between LDM and atomic electrons $f^{nl}_{\textrm{ion}} (k^\prime, q)$ with final momentum $k^\prime=\sqrt{2m_e E_{\textrm{ER}}}$ is the key parameter, which we take from~\cite{Essig:2012yx}.
The differential rates of the LDM-nucleus and LDM-electron scatterings calculated based on Eqs.~\ref{eq:nr} and ~\ref{eq:er} are shown in the top panels in Fig.~\ref{fig:signal_and_efficiency}.

% Data selection
This study utilizes data from the first commissioning run (run0) and the first scientific run (run1) of PandaX-4T, with durations of 95 and 164 calendar days~\cite{bo2025dark}, respectively.
Following the same methodology and event selection of the $^8$B solar neutrino search in Ref.~\cite{bo2024first}, we examine both paired events (containing both $S1$ and $S2$ signals) and unpaired ionization-only events (US2). 
This combined approach enables sensitivity across a wide energy range: approximately 0.33–3.0\,keV for NRs and 0.04–0.39\,keV for ERs. 
After removing high-rate live time and applying fiducial volume selection, the effective total exposures for the paired and US2 data are 1.20 and 1.04 tonne$\cdot$year, respectively. 
We employ various quality selections, including a boosted decision tree (BDT) method for the paired data, based on the signal amplitudes, pulse shapes, and PMT patterns. 
For the paired data, we require the $S1$ signal to be detected by two or three PMTs, while the US2 data must have the number of electrons for an $S2$ to be between four and eight detected electrons and no $S1$ signal detected by two or more PMTs in the event window. 
The total efficiencies, including contributions from the quality selections, the BDT cuts, the region-of-interest (ROI) cuts, and the signal reconstruction, as a function of NR and ER energies, are presented in the middle panels in Fig.\ref{fig:signal_and_efficiency}, and can also be found in the Supplementary Material~\cite{SupplementaryMaterial}.
More details of the data selection are given in Ref.~\cite{bo2024first}.
The efficiencies of paired data for ER are not given, since there are almost no expected LDM-electron scattering events in paired datasets due to the much lower light-to-charge ratio for the ER signals. 
This channel is not included in statistical analysis of LDM-electron interaction.

\begin{figure}[htp]
    \centering
    \includegraphics[width=0.90\columnwidth]{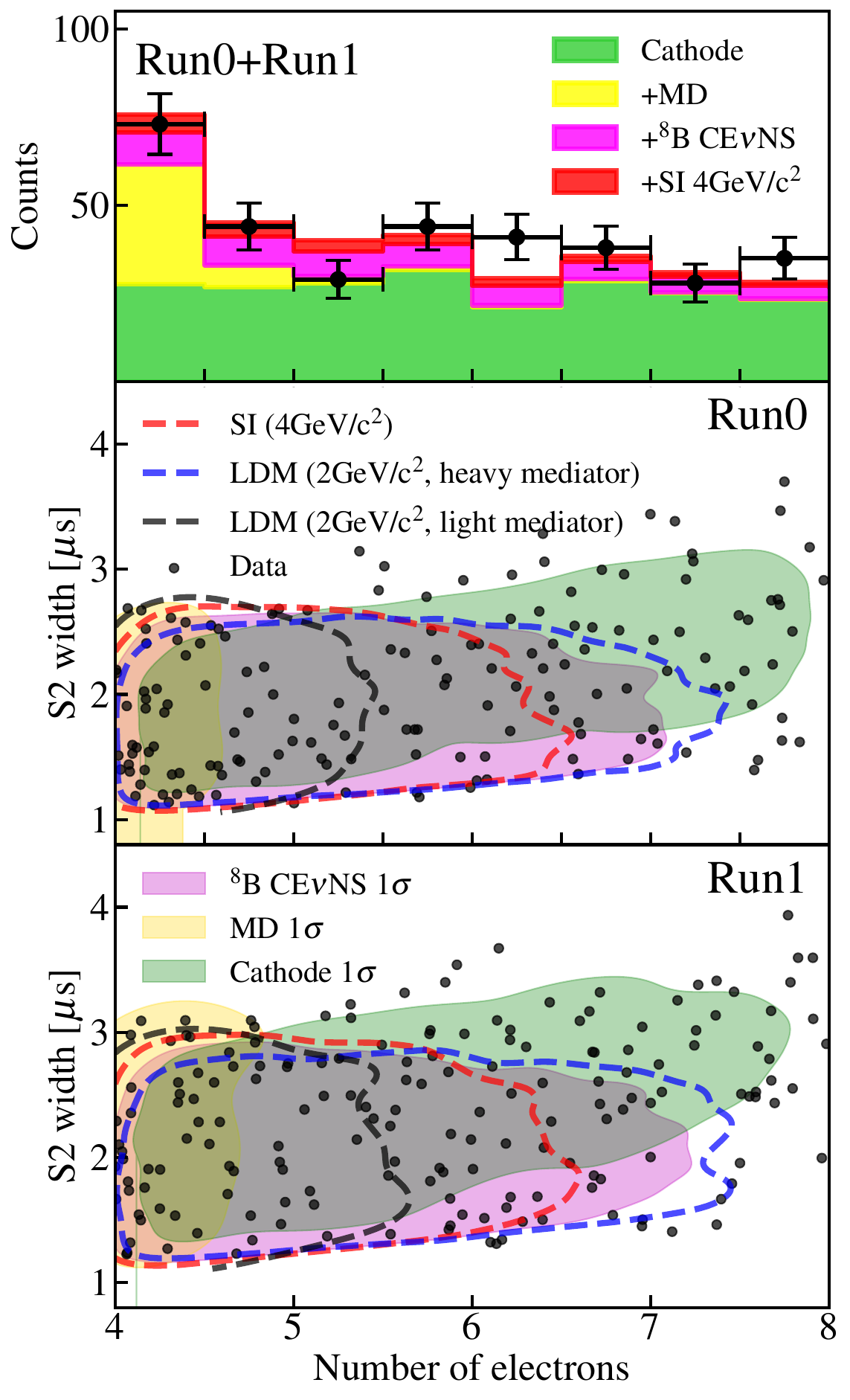}
    \caption{
    Top panel: US2 data and fitted (for a LDM of 4\,GeV/$c^2$) $S2$ spectra in number of electrons, run0 and run1 combined.
    % The one-dimensional $S2$ best-fit charge (number of electrons) spectrum in Run0 and Run1 combined for SI LDM with mass of 4\,GeV/c$^2$.
    The spectrum is stacked with best-fit contributions from the cathode (green), MD (yellow), $^8$B CE$\nu$NS (magenta), and LDM (red). 
    The black error bars give the data distributions.
    Middle and bottom panels: the US2 data(black dots) distribution of the $S2$ width versus $S2$ charge. Along with expected 1$\sigma$ contours of 4\,GeV/$c^2$ DM-nucleon spin-independent interaction(red dashed line), 2\,GeV/$c^2$ DM-electron interaction via heavy mediator(blue dashed line), and light mediator(black dashed line), ${}^8$B solar neutrino(magenta shadow), MD background(yellow shadow), and cathode background(green shadow). 
    }
    \label{fig:data_distribution}
\end{figure}

\begin{table*}[htp]
    \centering
    \begin{tabular}{cccccc}
    \hline\hline
    \multirow{2}{*}{Component}          &
    \multirow{2}{*}{Expectation}        &
    \multirow{2}{*}{Background-only}    &
    SI                                  &
    LDM-e (heavy mediator)              &
    LDM-e (light mediator)             \\
        &   &   &         
    4\,GeV/$c^2$                    &
    2\,GeV/$c^2$                    &
    2\,GeV/$c^2$                    \\
    \hline
    DM (paired)                                   &
    ···                                        &
    ···                                        &
    0.4$\pm$0.4                                     &
    ···                                     &
    ···                                     \\
    
    $^8$B CE$\nu$NS (paired)                &
    3.4$\pm$0.6                                     &
    2.6$\pm$0.6                                     &
    2.5$\pm$0.6                                     &
    ···                                     &
    ···                                     \\
    AC (paired)                             &
    2.5$\pm$0.5                                     &
    2.5$\pm$0.4                                     &
    2.5$\pm$0.4                                     &
    ···                                     &
    ···                                     \\
    \hline
    DM (US2)                                     &
    ···                                     &
    ···                                     &
    18$\pm$20                                     &
    58$\pm$34                                     &
    8$\pm$19                                     \\
    
    $^8$B CE$\nu$NS (US2)                &
    43$\pm$7                                     &
    50$\pm$6                                     &
    49$\pm$6                                     &
    42$\pm$6                                     &
    47$\pm$6                                     \\
    Cathode (US2)                              &
    204$\pm$32                                     &
    227$\pm$15                                     &
    212$\pm$19                                     &
    191$\pm$22                                     &
    225$\pm$17                                     \\
    MD (US2)                                      &
    45$\pm$5                                     &
    45$\pm$4                                     &
    45$\pm$4                                     &
    45$\pm$4                                     &
    45$\pm$4                                     \\
    \hline
    NR charge yield deviation                     &
    ···                                           &
    0.8$\pm$0.9$\sigma$                         &
    0.8$\pm$0.8$\sigma$                         &
    ···                         &
    ···                         \\
    ER charge yield deviation                     &
    ···                                           &
    0.0$\pm$1.0$\sigma$                         &
    ···                                           &
    0.2$\pm$1.0$\sigma$                         &
    0.0$\pm$1.0$\sigma$                         \\
    \hline\hline
    \end{tabular}
    \caption{The expected and best-fit events number of dark matter, ${}^8$B solar neutrino, AC background, cathode background and MD background in paired and US2 channels, assuming 4\,GeV/$c^2$ DM-nucleon spin-independent interaction and 2\,GeV/$c^2$ LDM-electron interaction via heavy mediator and light mediator. 
    The deviations of the best-fit charge yields with respect to their nominals for the NR and ER are given in the last two rows.}
    \label{tab:signal_bkg_numbers}
\end{table*}

% Signal & Background Model (focus on signal model)
The background models and their systematic uncertainties are primarily adopted from the previous solar neutrino analysis in Ref.~\cite{bo2024first}.
In the paired dataset, the dominant background arises from accidental pileup (AC) events, whereas the US2 dataset is primarily affected by three components: the cathode-originated background, the pileup of delayed electrons (MD background)~\footnote{
The acronym ``MD'' stands for micro-discharging since this kind of background was first observed during a dataset when discharging occurs quite frequently~\cite{PandaX:2022xqx}. 
}, and the coherent elastic neutrino-nucleus scattering induced by solar $^8$B neutrinos ($^8$B CE$\nu$NS).
The AC, cathode, and MD background models are constructed using data-driven approaches. 
As this is a search of the LDM signals on top of the solar neutrino CE$\nu$NS background, the CE$\nu$NS rate and spectrum are adopted from Ref.~\cite{ruppin2014complementarity} with the standard model~\cite{bahcall2005new}, instead of using our unconstrained fit to $^8$B flux in Ref.~\cite{bo2024first}.
The models have been tested with a sideband before unblinding as in Ref.~\cite{PandaX:2022xqx,PandaX:2022aac}.
% The rate and spectral shape of the $^8$B CE$\nu$NS are based on the calculations~\cite{ruppin2014complementarity} using the flux predicted by the Standard Solar Model~\cite{bahcall2005new, bahcall2005new}.
% The flux is chosen instead of our previous solar neutrino results~\cite{bo2024first} to ensure a conservative approach in constraining LDM signals.
Systematic uncertainties are assigned to the selection acceptance, BDT efficiency, MD background rate, cathode background rate, and signal response model, following the same methodology as Ref.~\cite{bo2024first}. 
The total background expectations for both the paired and US2 datasets, along with their corresponding uncertainties, are summarized in Table~\ref{tab:signal_bkg_numbers}.
The background models used in this work can be found in the Supplementary Material~\cite{SupplementaryMaterial}.

The signal response models for LDM and $^8$B CE$\nu$NS interactions in both the paired and US2 data are constructed using the PandaX-4T signal response framework (bottom panels in Fig.~\ref{fig:signal_and_efficiency})~\cite{luo2024signal}. 
The light and charge yields for NRs in liquid xenon, along with their associated uncertainties, are sourced from the NEST model~\cite{szydagis2022review}. 
For ERs induced by LDM-electron scattering, conservative light and charge yields are employed assuming a constant average energy of 13.8\,eV for producing one scintillation photon or ionization electron (similar to Ref.~\cite{PandaX:2022xqx}) in liquid xenon. 
% Alternatively, we also interpret the results using a more aggressive but robust light and charge yields from NEST~\cite{szydagis2022review}. 
For comparison, we also interpret the results using the charge yields from the Noble Element Simulation Technique (NEST) model~\cite{szydagis2022review}, leading to slightly more aggressive results.
% The light and charge yields for NRs and ERs are illustrated in the bottom panels of Fig.\ref{fig:signal_and_efficiency}.
% When interpreting the data into LDM-nucleus scattering, only the NR signal model uncertainty is assumed.
% But for the LDM-electron scattering, 
Both the NR and ER signal model uncertainties are considered in the likelihood analysis with the same treatment in Refs.~\cite{bo2024first, PandaX:2022xqx}.

% Data distribution and Statistical treatment
In the paired (US2) dataset, we observe one (158) and two (174) events in run0 and run1~\cite{PandaXUS2Data}, respectively. 
We conduct a binned profile likelihood ratio (PLR) analysis~\cite{cowan2011asymptotic} incorporating data from both runs for both paired and US2 datasets.
For the US2 data, the likelihood function is defined based on two discriminative variables: the signal charge ($\mathcal{Q}$) and pulse width ($\mathcal{W}$).
The joint ($\mathcal{W}$, $\mathcal{Q}$) distributions of the MD and cathode backgrounds are derived from controlled samples in the experimental data, while those for DM and $^8$B CE$\nu$NS signals are generated by the signal response model, assuming a uniform spatial distribution within the fiducial volume(FV).
Fig.~\ref{fig:data_distribution} displays the 1$\sigma$ confidence contours for both background components and DM signals with masses of 4\,GeV/$c^2$ (SI DM-n) and 2\,GeV/$c^2$ (LDM-e), alongside the observed US2 event distributions from run0 and run1.
Table~\ref{tab:signal_bkg_numbers} summarizes the best-fit rates for background components and DM signals under both the background-only and signal-plus-background (LDM-electron scattering) hypotheses.
% Since there are almost no expected LDM-electron scattering events in paired datasets, this channel is not included in statistical analysis of LDM-electron interaction.

% Results
\begin{figure*}
    \centering
    \includegraphics[width=0.98\textwidth]{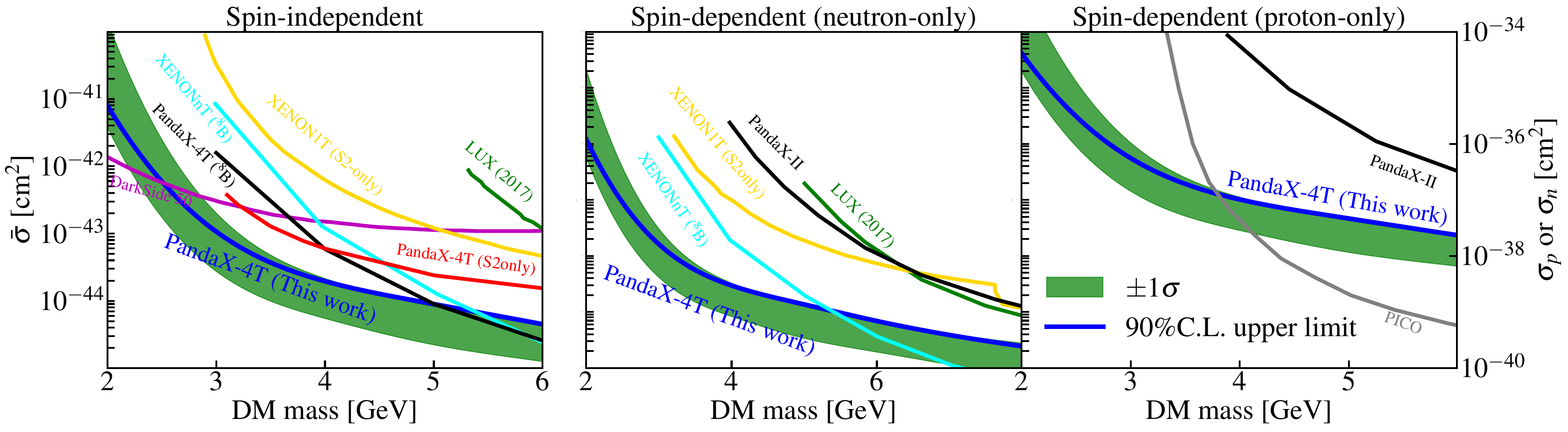}
    \caption{
    The 90\% CL exclusion limit(blue line) with 1$\sigma$ sensitivity(green band) on spin-independent DM-nucleon cross section(left panel), neutron-only(middle panel) and proton-only(right panel) spin-dependent DM-nucleon cross section, together with results from other work~\cite{PandaX:2022xqx, PandaX:2022aac, XENON:2024hup, XENON:2019gfn, DarkSide:2018kuk, LUX:2016ggv, PandaX-II:2018woa, PICO:2019vsc}.
    }
    \label{fig:nr_limits}
\end{figure*} 

\begin{figure}
    \centering
    \includegraphics[width=0.95\columnwidth]{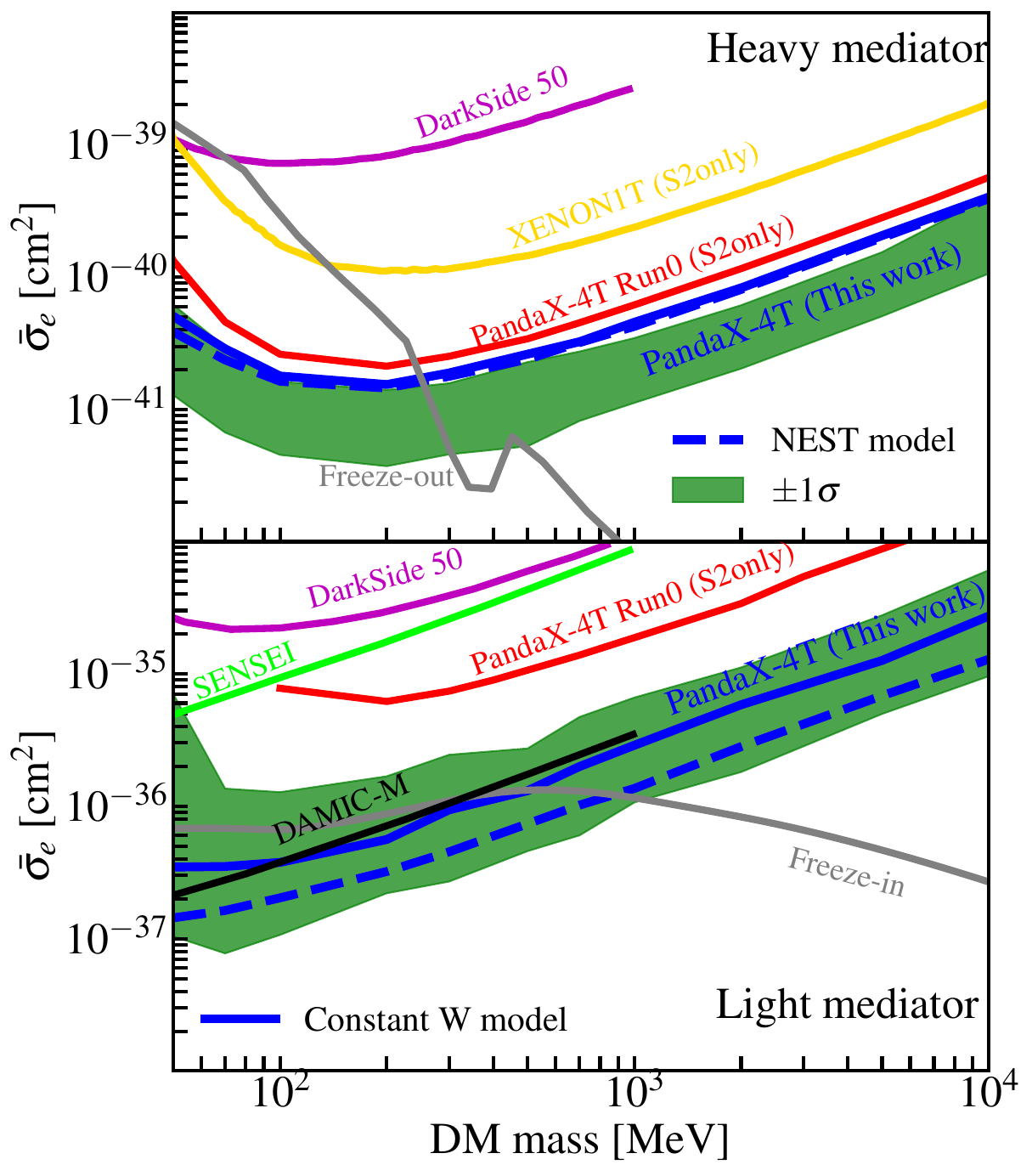}
    \caption{
    The 90\% CL exclusion limit(blue line) with 1$\sigma$ sensitivity(green band) on DM-electron cross section via heavy(top panel) and light mediator(bottom panel) using constant-W model. The aggressive results using NEST light and charge yields are also overlaid(blue dashed line), as well as results of other experiments~\cite{PandaX:2022xqx, XENON:2019gfn, DarkSide:2022knj, adari2025first, aggarwal2025probing}, and cosmological predictions from DM vector-portal freeze-in and freeze-out mechanisms~\cite{Essig:2015cda, bhattiprolu2024dark}.
    }
    \label{fig:er_limits}
\end{figure}

This analysis investigates five benchmark DM interaction models: 
SI DM-nucleon scattering, proton-only SD DM-nucleon scattering, neutron-only SD DM-nucleon scattering, and LDM-electron scatterings with heavy mediator and light mediator.
Figure~\ref{fig:nr_limits} presents the 90\% confidence level (CL) exclusion limits on the DM-nucleon scattering cross sections.
The results establish the most stringent limits for SI, SD (neutron-only), and SD (proton-only) DM-nucleon scatterings in the mass range of approximately [2.5, 5.0], [2.0, 5.3], and [2.0, 3.8]\,GeV/c$^2$, respectively.
The corresponding exclusion limits at 3\,GeV/$c^2$ are $1.1 \times 10^{-43}$, $1.6 \times 10^{-38}$, and $5.6 \times 10^{-37}$\,cm$^2$.
The enhanced exposure and reduced background in run1 US2 data yield significant improvements over previous PandaX-4T results~\cite{PandaX:2022xqx}.
The SI limits exhibit stronger upward fluctuations above $\sim$3.5\,GeV/$c^2$ because the harder energy spectrum populates the $S2$ region (5-8 electrons) where we observe a slight event excess. 
Below 3.5\,GeV/$c^2$, SI limits show no significant upward fluctuations as the spectra become dominated by MD background.
This trend also holds for the SD limits.
The LDM-electron cross section exclusion limits are improved by approximately 1.5 times for the heavy-mediator case, and 9.3 times for the light-mediator case, in mass range of $\sim$ [50, 10000]\,MeV/$c^2$.
% The improvement in light-mediator case is mainly because of different treatments in signal model: 
% ~\footnote{
% In previous work~\cite{PandaX:2022xqx}, the mean total number of created photons and electrons was conservatively set to a floored integer. 
% This treatment has a significant impact on the spectrum near our ROI threshold, particularly affecting searches for LDM–electron interactions mediated by a light mediator.
% }.
The improved sensitivity for light mediators stems from a refined signal model. 
Unlike the previous approach~\cite{PandaX:2022xqx}, which conservatively floored the mean number of photons and electrons to an integer, this work treats it as a continuous value, providing a more accurate spectrum near the ROI threshold critical for LDM–electron interactions mediated by a light mediator.
The limits for LDM with heavy mediators exhibit stronger upward fluctuations than those with light mediators, as heavy-mediator spectra predominantly populate the excess region (5-8 electrons), while light-mediator spectra primarily occupy the MD region (4-5 electrons).
% As a result, the expected number of DM-e interaction via light mediator is underestimated by about 3 times. This requirement is now removed, leading to a same degree of improvement on limits. 
In addition, due to the expected events gathering in the low-energy region where MD is dominant, greatly reduced MD uncertainty(from 130\%~\cite{PandaX:2022xqx} to 13\% for run0 and 16\% for run1) also improves the limits by around 2 times in light-mediator case.
The limit now has excluded portion of the cosmologically allowed parameter space from the cosmic microwave background measurements~\cite{Essig:2015cda,bhattiprolu2024dark}, in the mass range from about 50\,MeV/$c^2$ to 10\,GeV/$c^2$, which is shown in Fig.~\ref{fig:er_limits}.

% Summary & Discussion
In summary, we present a search for dark matter interactions using the complete run0 and run1 datasets from PandaX-4T, achieving world-leading sensitivity through a joint analysis of both scintillation-ionization paired events (1.20 tonne$\cdot$year exposure) and ionization-only events (1.04 tonne$\cdot$year exposure).
Our analysis establishes new exclusion limits at 90\% CL for: spin-independent DM-nucleon scattering in the mass range [2.5, 5.0]\,GeV/$c^2$, spin-dependent DM-nucleon scattering ([2.0, 5.3]\,GeV/$c^2$ for neutron-only coupling, and [2.0, 3.8]\,GeV/$c^2$ for proton-only coupling), and LDM-electron scattering (from 100\,MeV/$c^2$ to 10\,GeV/$c^2$).
The improved SI and SD limits, enabled by our lowered threshold through ionization-only events, significantly surpass previous results~\cite{PandaX:2022xqx, PandaX:2022aac, XENON:2024hup, XENON:2019gfn, DarkSide:2018kuk, LUX:2016ggv, PandaX-II:2018woa, PICO:2019vsc}.
For the  LDM model, our results constrain previously unexplored parameter space given by the cosmological predictions~\cite{Essig:2015cda, bhattiprolu2024dark}. 
These findings also demonstrate PandaX-4T's capability to probe new physics despite the neutrino CE$\nu$NS background.
The projected run2 dataset, with its anticipated reduction in background rates, is expected to further improve the sensitivity for the LDM searches.

% !TEX root = ../main.

%\section{Acknowledgement}
 
%the following are the grants for the whole collaboration, other personal grants can be added by the authors upon requests

$Acknowledgements$—This project is supported in part by grants from National Key R\&D Program of China (Nos. 2023YFA1606200, 2023YFA1606201), National Science Foundation of China (Nos. 12090060, 12090061, 12090063, 12325505, 12222505, 12275267, U23B2070), and by Office of Science and Technology, Shanghai Municipal Government (grant Nos. 22JC1410100, 21TQ1400218), Sichuan Province Innovative Talent Funding Project for Postdoctoral Fellows (No. BX202322), Sichuan Provincial Natural Science Foundation (No. 2024NSFSC1374). We thank for the support by the Fundamental Research Funds for the Central Universities. We also thank the sponsorship from the Chinese Academy of Sciences Center for Excellence in Particle Physics (CCEPP), Thomas and Linda Lau Family Foundation in Hong Kong, New Cornerstone Science Foundation, Tencent Foundation in China, and Yangyang Development Fund. Finally, we thank the CJPL administration and the Yalong River Hydropower Development Company Ltd. for indispensable logistical support and other help.

$Data\,availability$—The data that support the findings of this article are openly available\cite{PandaXUS2Data}.

%\bibliographystyle{apsrev4-2.bst}
%\bibliography{apssamp}

\end{document}